\documentclass[5p]{elsarticle}
\usepackage[T1]{fontenc}
\usepackage{lineno,hyperref}
\usepackage{amsfonts}
\usepackage{amssymb}
\usepackage{amsmath}
\usepackage{amsthm}
\usepackage{newlfont}
\usepackage{mathrsfs}
\usepackage{algorithm}
\usepackage{color}
\usepackage{graphicx}
\usepackage{epstopdf}
\usepackage{multirow}

\newcommand{\R}{\mathbb R}
\newcommand{\C}{\mathbb C}
\newcommand{\D}{\mathbb D}

\newcommand{\balancedVPhantom}[1]{
  $\mathsurround=0pt \vcenter{\hrule width0pt height #1}$\ignorespaces
}

\newtheorem{algor}{{\sc Algorithm}}[section]

\newtheoremstyle{itstyle}
  {\topsep}      
  {\topsep}      
  {\normalfont}  
  {0pt}          
  {\itshape \bf}     
  {.\ }          
  {0pt}          
  {\thmname{#1} \thmnumber{#2}}
\theoremstyle{itstyle}
\newtheorem{example}{\textit{Example}}

\def\betab{\begin{tabbing}
xxxx\=xxxx\=xxx\=xx\=xx\=xx\=xx\=xx\=xx\=xx\=xx\=xx\=xx\= \kill}
\def\entab{\end{tabbing}\vspace{-0.12in}}

\definecolor{darkviolet}{rgb}{0.58, 0.0, 0.83}

\makeatletter
\def\ps@pprintTitle{%
   \let\@oddhead\@empty
   \let\@evenhead\@empty
   \let\@oddfoot\@empty
   \let\@evenfoot\@oddfoot
}
\makeatother

\begin{document}

\begin{frontmatter}

\title{FEAST Eigensolver for Nonlinear Eigenvalue Problems}

\author[BG]{Brendan Gavin}
\address[BG]{Department of Electrical and Computer Engineering, University of Massachusetts, Amherst, MA 01003, USA, \texttt{bgavin@ecs.umass.edu}}

\author[AM]{Agnieszka Mi\k{e}dlar}
\address[AM]{Department of Mathematics, University of Kansas, 1460 Jayhawk Blvd., Lawrence, KS 66045-7594, USA, \texttt{amiedlar@ku.edu}}

\author[EP]{Eric Polizzi}
\address[EP]{Department of Electrical and Computer Engineering, Department of Mathematics and Statistics, University of Massachusetts, Amherst, MA 01003, USA, \texttt{polizzi@ecs.umass.edu}}


%
%

\begin{abstract}

The linear FEAST algorithm is a method for solving linear eigenvalue problems. It uses complex contour integration to calculate the eigenvectors whose eigenvalues that are located inside some user-defined region in the complex plane. This makes it possible to parallelize the process of solving eigenvalue problems by simply dividing the complex plane into a collection of disjoint regions and calculating the eigenpairs in each region independently of the eigenpairs in the other regions. In this paper we present a generalization of the linear FEAST algorithm that can be used to solve nonlinear eigenvalue problems. Like its linear progenitor, the nonlinear FEAST algorithm can be used to solve nonlinear eigenvalue problems for the eigenpairs whose eigenvalues lie in a user-defined region in the complex plane, thereby allowing for the calculation of large numbers of eigenpairs in parallel. We describe the nonlinear FEAST algorithm, and use several physically-motivated examples to demonstrate its properties.
\end{abstract}


\end{frontmatter}


\section{Introduction}


\noindent
The nonlinear eigenvalue problem (NLEVP) consists of finding vectors $x\in {\mathbb C}^{n}$ and scalars $\lambda \in {\mathbb C}$ that satisfy

\begin{equation}
  \label{eqn:nlevp}
T(\lambda)x=0,
\end{equation}

\noindent where $T(\lambda)\in {\mathbb C}^{n\times n}$ is some matrix-valued function of $\lambda$, which we call the eigenvector residual function. The linear generalized eigenvalue problem (GEP) is a special case of (\ref{eqn:nlevp}), with

\begin{equation}
T(\lambda)=\lambda B-A,\ \ \ A,B\in {\mathbb C}^{n\times n}.
\end{equation}

\noindent  Another well-known example of a NLEVP is a quadratic eigenvalue problem

\begin{equation}
T(\lambda)=\lambda^2 A_2 +\lambda A_1 + A_0,\ \ \ A_2,A_1,A_0\in {\mathbb C}^{n\times n}.
\end{equation}

\noindent Quadratic eigenvalue problems can arise, for example, in models of physical systems undergoing damped oscillations, e.g.,~\cite{HolGL05,LiC15}.
In general $T(\lambda)$ can take a variety of different forms, depending on the physical model. Higher-degree polynomials are possible, as are non-polynomial functions of $\lambda$, e.g. ~\cite{MehW02,Kau06,Sol06,Lia07}.

The usefulness of any physical model that produces the problem (\ref{eqn:nlevp})
is limited by the ease with which that problem can be solved.
General nonlinear eigenvalue problems carry with them some unique challenges, e.g.
their eigenvectors do not form a basis for $\mathbb C^n$, and the particular form of $T(\lambda)$ can have an impact on which solution methods are most effective.
However, one of the core challenges for nonlinear eigenvalue problems is the same as for linear eigenvalue problems:
solution methods that are effective for small values of $n$ do not necessarily scale well to very large-dimensional problems,
where one is usually interested in finding a small number of specific, physically-important eigenvalue-eigenvector pairs.

Most methods for solving such large-dimensional problems require generating
an initial guess that is close enough to the desired eigenpairs to ensure convergence,
as well as a method for calculating successive eigenpairs whose eigenvalues are close to each other,
but without converging repeatedly to the same eigenpair \cite{Vos14}.
These challenges are exacerbated in the situation that large numbers of eigenpairs are desired.
The difficulty of calculating large numbers of eigenpairs is compounded further for NLEVP algorithms that work by approximating the desired eigenvectors in some subspace of dimension $m\ll n$;
in this case, calculating large numbers of eigenpairs requires a large value of $m$, leading to bad scaling for the solution of the reduced-dimension eigenvalue problems that are solved inside that subspace. For linear eigenvalue problems the algorithmic complexity of solving the reduced dimension problem of dimension $m$ is $O(m^3)$, and solving nonlinear eigenvalue problems always requires an amount of work that is at least equal to solving a single linear eigenvalue problem; keeping the dimension $m$ of the reduced-size problem small is thus of paramount importance for ensuring good scaling to larger problem sizes.

A class of algorithms that address these challenges is contour integral methods, e.g.,~\cite{AsaSTIK09,AsaSTIK10,Bey12}.
Contour integral methods exploit the Cauchy integral formula in order to calculate only the eigenpairs whose eigenvalues lie in a specific, user-defined region of the complex plane.
In doing so they eliminate the need to calculate initial guesses and separate closely-positioned eigenvalues for the original, large problem of dimension $n$.
Moreover, contour integral approaches allow for the efficient, parallel calculations of large numbers of eigenpairs by
dividing the region of interest in the complex plane into disjoint subregions, solving for the eigenvalues in each subregion independently of the eigenvalues elsewhere.

The linear FEAST algorithm is an example of a contour integral method for linear eigenvalue problems.
It has the distinction of being able to solve large, generalized linear eigenvalue problems robustly with excellent parallel scalability.
In the following sections, we describe a generalization of the linear FEAST algorithm that can be applied to solving nonlinear eigenvalue problems.
We describe the resulting nonlinear FEAST (NLFEAST) algorithm and solve some physically-motivated model problems to illustrate its properties.

\section{Nonlinear Eigenvalue Problems}
\noindent
Given a non-empty open set $\D \subseteq \C$ and a matrix-valued function $T: \D \rightarrow \C^{n \times n}$,
we consider a \textit{nonlinear eigenvalue problem}
(NLEVP):
\emph{Find a scalar $\lambda \in \D$ and nonzero vectors $x \in \C^n$ and $y \in \C^n$ such that
\begin{equation}
\label{eq:NLEVP}
T(\lambda)x = 0, \quad y^*T(\lambda) = 0.
\end{equation}
}
In this paper, we discuss only the nonlinear dependence of $T$ on the eigenparameter $\lambda$,
leaving the nonlinear dependence on eigenvectors and multi-parameters outside of our focus.
Analogously to the linear case, $\lambda \in \D$ is called a \textit{eigenvalue}
and $x, y$ the corresponding right and left \textit{eigenvectors} of $T$, respectively.
We call the set of all eigenvalues $\lambda$ of $T$ the
\textit{spectrum} of $T$ and denote i by $\sigma(T)$, i.e.,
\[
\sigma(T) := \Big\{ \lambda \in \D: \mbox{det}(T(\lambda)) = 0 \Big\}.
\]
We refer to $(\lambda,x,y)$ as an \textit{eigentriple} of $T$, and either $(\lambda, y)$ or $(\lambda, x)$ as an \textit{eigenpair}.
Although for the choice $T(\lambda) = \lambda I - A$ and $T(\lambda) = \lambda B - A$,
\eqref{eq:NLEVP} reduces to the standard eigenvalue problem $Ax = \lambda x$, and the generalized eigenvalue problem
$Ax = \lambda B x$, respectively, the characteristics of a general nonlinear eigenvalue problem differ significantly from those of its
linear counterparts, e.g. $T$ (even if regular, i.e., det$(T(\lambda)) \neq 0$) may have infinitely many eigenvalues, the eigenvectors belonging to distinct eigenvalues do not
have to be linearly independent, and the algebraic multiplicity of an isolated eigenvalue, although finite, is not bounded by the problem size $n$~\cite{TisG17}.
All these particularities make general nonlinear eigenvalue problems much harder to solve.
For a more detailed description we refer the readers to the survey papers by
Tisseur and Meerbergen~\cite{TisM01} for quadratic eigenvalue problems,
Mackey, Mackey and Tisseur~\cite{MacMT15} for general polynomial eigenvalue problems,
and by Voss~\cite{Vos14}, and Tisseur and G\"{u}ttel~\cite{TisG17}
for general nonlinear eigenvalue problems. A variety of applications in science and
engineering, e.g., dynamic analysis of structures, vibrations of fluid-solid structures,
the electronic behavior of quantum dots are discussed by Mehrmann and Voss~\cite{MehV04}.

For the sake of simplictly, our main focus in this paper is on a class of square \textit{polynomial eigenvalue problems} (PEPs):
\emph{Find a scalar $\lambda \in \D \subset \C$ and nonzero vectors $x,y \in \C^n$ such that
\begin{equation}
\label{eq:PEP}
P(\lambda) x = 0, \quad y^*P(\lambda) = 0,
\end{equation}
}
where $P(\lambda)$ is an $n \times n$ matrix polynomial
\begin{equation}
  \label{eqn:polynomialeig}
P(\lambda) = \sum\limits_{i = 0}^k \lambda^i A_i, \quad A_i \in \C^{n \times n}.
\end{equation}
If $P$ is regular, i.e., det$(P) \neq 0$, \eqref{eq:PEP} has $r$ finite eigenvalues (roots of det$(P)=0$) and $kn - r$ infinite eigenvalues.
Using linearization, it is easy to see that the $n \times n$ matrix polynomial $P(\lambda)$ of degree $k$ has $kn$ eigenvalues (finite or infinite)
and up to $kn$ right and $kn$ left associated eigenvectors. As mentioned before, even in the case of distinct eigenvalues,
the associated eigenvectors are not necessarily linearly independent.
%
%

For example, let us consider the \textit{quadratic eigenvalue problem} (QEP):
\emph{Find a scalar $\lambda \in \D \subset \C$ and nonzero vectors $x,y \in \C^n$ such that}
\begin{align}
\label{eq:QEP}
Q(\lambda)x = \big( \lambda^2 A_2 + \lambda A_1 + A_0) x = 0,\\
y^*Q(\lambda) =  y^*\Big( \lambda^2 A_2 + \lambda A_1 + A_0) = 0,
\end{align}
with $n \times n$ complex matrices $A_2, A_1$, and $A_0$.
Then the eigenvalues of $Q(\lambda)$ are equal to those of the $2n\times 2n$
companion linearization~\cite[Ch.~1]{GohLR09}, e.g., 
\begin{equation}
\nonumber
L(\lambda) =
  \lambda\begin{bmatrix}
  A_2& O \\ O&I
  \end{bmatrix} + \begin{bmatrix}
  A_1& A_0\\-I&O
  \end{bmatrix} \quad \mbox{(the first companion form) }
\end{equation}
or
\begin{equation}
\nonumber
L(\lambda) =
  \lambda\begin{bmatrix}
  A_2& O \\ O&-A_0
  \end{bmatrix} + \begin{bmatrix}
  A_1& A_0\\A_0&O
  \end{bmatrix}.
\end{equation}
The right eigenvectors of $L$ are
$\begin{bmatrix}
\lambda x  \\ x
\end{bmatrix}$, where $x$ are the right eigenvectors of $Q(\lambda)$.
Similarly, the left eigenvectors of $L$ are
$
\begin{bmatrix}
 y \\ (\lambda A_2+A_1)^*y
\end{bmatrix}$,
where $y$ are the left eigenvectors of $Q(\lambda)$.
%
%

\section{The FEAST Algorithm}

\noindent
The original FEAST algorithm, introduced for solving the generalized eigenvalue problems, i.e. $T(\lambda) =  \lambda B - A$~\cite{Pol09},
is a subspace iteration method that uses the Rayleigh-Ritz projection and an approximate spectral projector as a filter.
%
%
FEAST calculates the eigenpairs whose eigenvalues lie in a specific, user-defined region in the complex plane.
It can therefore be used to calculate a large number of eigenpairs in parallel by dividing the complex plane
into a collection of disjoint regions and solving for the eigenpairs in each region independently of the eigenpairs in other regions.

FEAST selects the eigenvectors and eigenvalues of interest by first projecting any random initial subspace $X^{(0)}$
onto the subspace spanned by the eigenvectors of interest, and then using the Rayleigh-Ritz procedure in this subspace to extract eigenvalue/eigenvector approximations.
Projection onto the subspace of interest is accomplished by using complex contour integration, i.e.
\begin{equation}
\label{eqn:feast_int}
Q=\rho(A,B)X^{(0)}=\frac{1}{2\pi i}\oint _{\cal C} (zB-A)^{-1}Bdz\ X^{(0)},
\end{equation}

\noindent where $Q$ is the new, projected subspace, and the integration is performed in a closed contour $\cal C$ around the region of the complex plane where we would like to find eigenvalues.
If the integration is performed exactly then the filter $\rho(A,B)$ becomes a spectral projector with $\rho(A,B)=XY^HB$,
where $X$ and $Y$ are the exact left and right eigenvector subspaces associated with the eigenvalues of interest (i.e. the eigenvalues that are located within the closed curve $\cal C$).
In practice, however, it is impossible to perform the integration in (\ref{eqn:feast_int}) exactly, so instead it is approximated by using a quadrature rule.
The subspace that is generated by the practical FEAST algorithm is then

\begin{equation}
\label{eqn:feast_sum}
Q=\sum _{j=1}^{n_c} \omega_j(z_jB-A)^{-1}BX^{(0)} = \sum _{j}^{n_c} \omega _j Q_j,
\end{equation}

\noindent where $z_j$ and $\omega_j$ are integration nodes and weights, respectively, and the summation is performed by solving linear systems $(z_jB-A)Q_j=BX^{(0)}$. Because each of these linear systems is independent of the others, they can all be solved simultaneously in parallel.
FEAST iteratively refines the estimates for the eigenvectors of interest by repeatedly applying the summation (\ref{eqn:feast_sum}), orthogonalizing the resulting subspace $Q$ and using it in the Rayleigh-Ritz procedure.

In addition to parallelizing the solution of eigenvalue problems by dividing up the eigenspectrum,
FEAST has the additional benefit of being able to systematically improve its rate of convergence.
The rate of convergence of the FEAST algorithm is related to the accuracy of the approximation for the integral (\ref{eqn:feast_int}) \cite{TanP14}.
If that integral is evaluated exactly, then the exact solution to the eigenvalue problem is found in a single iteration.
Approximating (\ref{eqn:feast_int}) by using (\ref{eqn:feast_sum}) generally means that the solution to the eigenvalue problem
will be found in some number of iterations larger than one;
reducing the number of required iterations is as simple as increasing the number of terms in the quadrature rule summation
in (\ref{eqn:feast_sum}), which in turn can be accomplished by solving a larger number of linear systems in parallel.
The speed with which an eigenvalue problem can be solved by using FEAST
is thus limited only by the amount of parallel processing power that is available.

The FEAST numerical library package was first released under the free BSD license in September 2009 (v1.0) to address the Hermitian linear eigenvalue problems.
The current FEAST version (v.3) released in June 2015 allows to solve the non-Hermitian linear eigenvalue problems, see~\cite{KesPT16}.

In the following section, we discuss an extension of the FEAST algorithm to tackle  nonlinear eigenvalue problems.
Our goal is to solve NLEVPs by taking advantage of the attractive features of the FEAST framework:
we would like to solve for eigenpairs whose eigenvalues are in a user-defined region of the complex plane, by iteratively refining a fixed-dimension
subspace using contour integration in such a way that the rate of convergence can be enhanced by using parallel computing to improve the accuracy of the contour integration.

%

\section{FEAST for Nonlinear Eigenvalue Problems}

\subsection{Previous and Related Work}

Other researchers have previously investigated the use of contour integration (using Cauchy's integral in particular) in the
complex plane for solving nonlinear eigenvalue problems.

In a series of papers Asakura at al. introduced the nonlinear variants of the Sakurai-Sugiura (SS) method with block Hankel matrices (SS-H method)
for the polynomial eigenvalue problems~\cite{AsaSTIK09} and eigenvalue problems of analytic matrix functions $T(\lambda)$~\cite{AsaSTIK10}.
Although they are cost efficient and highly scalable, the accuracy of obtained solutions is relatively low.
Almost at the same time Beyn~\cite{Bey12} introduced a highly accurate algorithm that uses the zeroth and first-order moments matrices
to reduce an NLEVP with $m \ll n$ eigenvalues inside $\cal C$ to a linear eigenvalue problem of dimension $m$.
The main idea of Beyn's integral method is to probe a Jordan decomposition of the $m \times m$ matrix following the Keldysh's theorem~\cite[Theorem 1.6.5]{MennickenMoller},
which is conceptually very simple but known to be highly sensitive to perturbations. Moreover, since the value of parameter $m$ (the number of linearly independent eigenvectors)
is not known in advance, the practical realization of Beyn's algorithm requires various adaptations which makes its overall computational cost relatively high.
%
%

Recently, Yokota and Sakurai~\cite{YokS13} addressed the problem of low accuracy in the nonlinear SS-H method.
Their method and the nonlinear variant of the Sakurai-Sugiura method with block Hankel matrices (SS-H)~\cite{AsaSTIK09}
use the same contour integrals of the SS-H method,
however, they differ in the way the approximate eigenpairs are extracted from the underlying subspaces.
The method of Yokota and Sakurai is a projection-based method which uses the Rayleigh-Ritz procedure to obtain the approximate
eigenpairs from a subspace. It does not require any fixed point iterations and
gives better accuracy than the methods of Asakura et al.~\cite{AsaSTIK09} and Beyn~\cite{Bey12}.

Both the various SS methods and the Beyn method use the moments of the Cauchy integral of $T^{-1}(\lambda)$, i.e.,

\begin{equation}
  \label{eqn:cauchymoments}
\frac{1}{2\pi i} \oint _{\mathcal C} z^kT^{-1}(z)dz,\ \ k\ge 0.
\end{equation}

\noindent Beyn's method uses the $k=0$ and $k=1$ moments, and the SS-type methods use as many moments as are necessary to reach convergence. 
The original FEAST algorithm uses only the $k=0$ moment, which, in the case when $T(\lambda)$ is linear, turns out to be a spectral projector 
associated with the eigenvalues of interest. Unfortunately, subspace iteration using only the $k=0$ moment in (\ref{eqn:cauchymoments}) does not work in the case of the nonlinear $T(\lambda)$. 
When $T(\lambda)$ is nonlinear, taking an initial set of approximate eigenvectors $X^{(0)}$ and refining it using the quadrature approximation of the $k = 0$
moment of (\ref{eqn:cauchymoments}), as is done in the linear FEAST algorithm, does not bring the resulting subspace closer to the desired eigenspace.
Although Beyn's method resolves this issue by continuously refining the accuracy of
the contour integration for a single multiplication, it requires performing a new matrix factorization for each
newly added quadrature node. The SS-type methods, on the other hand, refine their solutions by increasing the dimension of their search subspace by calculating additional moments of (\ref{eqn:cauchymoments})
using the same quadrature rule each time; the dimension of the search subspace is increased iteratively until the desired solution is sufficiently accurate.

We propose, instead, to modify the contour integral (\ref{eqn:cauchymoments}) 
such that the subspace iteration framework of the original FEAST algorithm can be used, thereby making it possible to solve nonlinear eigenvalue problems using a fixed-dimension subspace that is refined by solving linear systems at a constant number of fixed shifts in the complex plane.

\subsection{The Nonlinear FEAST Algorithm}

To develop a nonlinear FEAST algorithm that can use contour integration to iteratively refine a
subspace of a fixed dimension by using fixed-location shifts in the complex plane, we propose to study the
following (modified) form of the contour integral

\begin{equation}
\label{eqn:nlfeast_int1}
\frac{1}{2\pi i}\oint _{\cal C} \frac{I-T^{-1}(z)T(\lambda)}{z-\lambda} dz,
\end{equation}

\noindent where $\lambda$ is the current Ritz estimation 
of an eigenvalue that is inside the contour $\mathcal C$. For the implementation of nonlinear FEAST,
we use a block version of (\ref{eqn:nlfeast_int1}) in order to generate a refined a subspace from an initial set of approximate eigenvectors $X^{(0)}$ of a  nonlinear eigenvalue problem, i.e.,

\begin{equation}
\label{eqn:nlfeast_int}
Q=\frac{1}{2\pi i}\oint _{\cal C}  (X^{(0)}-T^{-1}(z)T(X^{(0)},\Lambda^{(0)}))(zI-\Lambda^{(0)})^{-1} dz,
\end{equation}

\noindent where $T(X^{(0)},\Lambda^{(0)})$ is the block eigenvector residual function for the initial estimate for the eigenvectors $X^{(0)}$, and the diagonal matrix of the corresponding Ritz values is $\Lambda^{(0)}$. 
For example, for the polynomial eigenvalue problem in equation (\ref{eqn:polynomialeig}), the block form of the residual function is

\begin{equation}
T(X,\Lambda) = \sum\limits_{i = 0}^k  A_iX\Lambda^{i}.
\end{equation}

\noindent
The nonlinear FEAST algorithm follows the same essential steps as the linear FEAST algorithm:
a subspace is formed by refining an initial set of estimated eigenvectors $X^{(0)}$ by using the contour integration in (\ref{eqn:nlfeast_int}),
then a new set of approximate eigenvectors is found in the resulting subspace $Q$ by using the Rayleigh-Ritz
procedure to solve a projected eigenvalue problem (which is still nonlinear, but with a substantially smaller dimension).
As in the linear case we use a numerical integration scheme to evaluate the integral in
(\ref{eqn:nlfeast_int}), and we solve a linear system whenever a multiplication by $T^{-1}(z)$ is required.

The contour integral in (\ref{eqn:nlfeast_int}) is mathematically equivalent to
(\ref{eqn:feast_int}) for the linear, generalized eigenvalue problem. However, the two are different when $T(\lambda)$ depends nonlinearly on $\lambda$.
The relationship between the contour integrals for linear and nonlinear FEAST is much like the relationship between the shift-and-invert iteration
for the linear eigenvalue problem and the residual inverse iteration for the nonlinear eigenvalue problem.
The residual inverse iteration~\cite{Neu85} is a modification of the shift-and-invert iteration which allows it to be applied to
to nonlinear problems while using a constant shift $\sigma \in \C$. The original FEAST algorithm can be interpreted as a generalization of the
shift-and-invert iteration that uses contour integration in order to efficiently use multiple shifts simultaneously.
We note that the nonlinear FEAST algorithm, in turn, can be interpreted as a generalization of residual inverse iteration that uses contour
integration in order to efficiently use multiple shifts simultaneously. 

In this paper, we will present the algorithmic framework and various example applications of the NLFEAST algorithm,
leaving the mathematical details to a forthcoming paper. The full NLFEAST algorithm is described in Algorithm \ref{alg:nlfeast}.

The Rayleigh-Ritz step in the nonlinear FEAST works the same as in its linear counterpart:
the residual function $T(\lambda)$ is projected onto the subspace $Q$,
and the resulting nonlinear eigenvalue problem of reduced-dimension
is solved using any suitable method, here we use a simple linearization.
The resulting Ritz values and associated Ritz vectors are used as the new estimates of the desired eigenpairs.

Careful attention must be paid while selecting the desired eigenpairs from the solutions of the nonlinear eigenvalue problem of reduced-dimension.
In many cases, such as those considered in this paper,
it is possible to find all the solutions of the reduced-dimension eigenvalue problem; for the nonlinear FEAST algorithm, however,
we want the Rayleigh-Ritz procedure to return only a number of eigenpairs that is equal to the dimension of the FEAST subspace $Q$.
The desired eigenpairs are thus selected by choosing those whose eigenvalues are closest to being inside the FEAST
search contour $\cal C$. This heuristic appears to ensure convergence to only the eigenpairs whose eigenvalues are inside the region of
interest in the complex plane.

The subspace that is used for the Rayleigh-Ritz procedure, i.e.,
the one that is generated by performing the contour integration, should be orthonormalized,
for example by using the $QR$ decomposition. Orthonormalization improves the numerical stability of the FEAST iterations
by preventing the norms of the desired eigenvectors from diverging and by preventing the Rayleigh-Ritz procedure from
producing spurious eigenpairs, i.e., the Ritz pairs that do not correspond to any eigenpairs of the full-size eigenvalue problem.

\begin{algorithm}[htp]

  \medskip
\noindent \begin{minipage}{\linewidth}

{\footnotesize
\noindent {\bf Start with:}
\begin{itemize}
\item Matrix function $T(\lambda) \in \mathbb{C}^{n \times n}, \ \ \lambda \in \D \subset \C$
\item Initial (possibly random) guess $X^{(0)} \in \mathbb{C}^{n\times m_0}$ for the search subspace (initial set of estimated eigenvectors)
\item Closed contour ${\cal C}$, inside of which fewer than $m_0$ eigenvalues are expected to be found
\item Set of $n_c$ quadrature nodes and weights $(z_j, \omega_j)$ for performing numerical integration over the contour $\mathcal C$

\end{itemize}
\medskip
{\bf Step 0.}  Set the search subspace $Q=X^{(0)}$, orthonormalize column vectors of $Q$

\medskip

\noindent {\bf For each} iteration $i$:
\medskip
\begin{description}
\setlength{\itemsep}{3pt}

\item[Step 1.] Solve the projected nonlinear eigenvalue problem
\begin{equation}
Q^HT(\lambda)Qy=0,
\end{equation}

\noindent for the approximate eigenpairs $(\lambda,Qy)$.

\item[Step 2.] Select the $m_0$ approximate eigenpairs $(\lambda_i,Qy_i)$ whose eigenvalues $\lambda$ are closest to the interior of the contour $\cal C$, and store these as an approximate invariant pair $(\Lambda, X)$

\begin{equation}
\Lambda= \mbox{diag}(\lambda_1, \lambda_2, ...,\lambda _{m_0}),\ \ \ X=[Qy_1,Qy_2,...,Qy_{m_0}].
\end{equation}

\item[Step 3.] If $||T(\lambda_i)Qy_i|| \leq \varepsilon$ for all $\lambda_i$ inside $\mathcal C$ \textbf{STOP}; otherwise continue.

\item[Step 4.] Update the search subspace by performing the numerical integration

\begin{equation}
Q=\frac{1}{2\pi i}\oint\limits_{\cal C}\left(X-T^{-1}(z)T(X,\Lambda)\right)(zI-\Lambda)^{-1}dz,
\end{equation}

\noindent using a quadrature rule and solving linear systems:

\begin{equation}
Q=\sum _{j=1}^{n_c} \omega_j\left(X-T^{-1}(z_j)T(X,\Lambda)\right)(z_jI-\Lambda)^{-1}.
\end{equation}

\item[Step 5.] Orthonormalize the column vectors of the new search subspace $Q$ (using e.g. the $QR$ decomposition), and go to {\bf Step 1.}

\end{description}
}
\end{minipage}
\caption{\label{algo:nlevpfeast} NLFEAST for NLEVPs $T(\lambda)x=0$}\label{alg:nlfeast}

\end{algorithm}

In the following section we describe the results of using Algorithm \ref{algo:nlevpfeast} to solve several example nonlinear eigenvalue problems.


\section{Numerical Examples}

\noindent
In this section, we illustrate the behavior of the nonlinear FEAST algorithm on several
polynomial eigenvalue problems of practical relevance.

\begin{example}{\it A nonoverdamped mass-spring system}
\label{Exp1}

\smallskip
\noindent

Let us consider the Hermitian (self-adjoint) quadratic eigenvalue problem (HQEP), i.e., $T(\lambda) = T(\bar \lambda)^H$,
associated with the mass-spring system discussed in~\cite[\S3.9]{TisM01}, i.e.,

\begin{equation}
\nonumber
T(\lambda)x = \big(\lambda^2A_2 + \lambda A_1 + A_0\big)x = 0,
\end{equation}
with \[A_2 = I_n, \quad A_1 = \tau \begin{bmatrix}
                           3 & -1 &  0 & 0 & \ldots & 0 \\
                          -1 &  3 & -1 &  0 & \ldots & 0 \\
                           0 & -1 &  3 & -1  & \ldots & 0 \\
                           \vdots & \vdots & \vdots & \ddots & \vdots & \vdots\\
                           0 & \ldots & 0 & -1 & 3 & -1 \\
                           0 & \ldots & 0 & 0 & -1 & 3
                          \end{bmatrix},\]
                          and
\[A_0 = \kappa \begin{bmatrix}
                           3 & -1 &  0 & 0 & \ldots & 0 \\
                          -1 &  3 & -1 &  0 & \ldots & 0 \\
                           0 & -1 &  3 & -1  & \ldots & 0 \\
                           \vdots & \vdots & \vdots & \ddots & \vdots & \vdots\\
                           0 & \ldots & 0 & -1 & 3 & -1 \\
                           0 & \ldots & 0 & 0 & -1 & 3
                          \end{bmatrix}.
                          \]
The eigenvalues of a Hermitian $T(\lambda)$ are either real or they come in complex conjugate pairs $(\lambda, \bar \lambda)$.
Since matrices $A_2,A_1$ and $A_0$ are real, the right and left eigenvectors coincide.

Following~\cite[\S4.2]{LiC15}, we first study the nonoverdamped system.
We set $n=1000$ and choose $\tau = 0.6202$, $\kappa= 0.4807$. All the eigenvalues of $Q$ are plotted in Figure~\ref{fig:nonALL}. 
One possible problem of interest is to calculate the eigenvectors whose eigenvalues have no imaginary part; the exponentially increasing 
or decaying solutions to the original equations of motion for the mass-spring system are linear combinations of these eigenvectors.
In this case, we seek the $(20)$ real eigenvalues lying within the interval $(-1.6, -1.5)$, see Figure~\ref{fig:nonReal}.

We use the nonlinear FEAST approach presented in Algorithm 1 to compute approximations of those $20$ real eigenvalues. The contour is chosen as
an ellipse centered at the midpoint of the interval $(-1.6,-1.5)$ with radius $r_a = 0.05$ on the real and $r_b = 0.0035$ on the imaginary axis using
$n_c = 16$ integration nodes, and the convergence criteria for the residuum is $tol = 10^{-10}$. With the subspace of size $m_0 = 22$,
the nonlinear FEAST algorithm finds all $m = 20$ eigenvalues within the interval in three FEAST iterations, see Figure~\ref{fig:nonReal}.
The eigenvalue approximations obtained using the linearization approach and the nonlinear FEAST algorithm
are listed in Table~\ref{tab:nonoverdamped}.
For completeness, we also list all the final residuals, i.e., {$\|r_i\|_2 = \| T(\lambda _i)x_i \|_2$}.

Table \ref{tab:interval} shows the number of NLFEAST iterations that are required for convergence 
when the search interval $(a,b)$ is enlarged. When the number of quadrature nodes $n_c$ 
is kept constant at $16$, as it is here,
then enlarging the search interval has the same effects as placing the quadrature nodes farther away from the eigenvalues of interest,  making the contour integration itself less accurate.
Modest changes to the size of the interval result in only a slightly larger amount of work being required for convergence; larger changes in the interval size tend to require 
proportionally more NLFEAST iterations. In practice, decreased performance due to the size of the search interval relative to the distribution 
of the desired eigenvalues can be mitigated simply by using a larger number of quadrature nodes which, in turn, requires only that one use more parallel processing power in order to solve more linear systems simultaneously.


\begin{figure}[tbh!]
\begin{center}
    \textbf{All Eigenvalues of the \\ Nonoverdamped Problem}\\
\includegraphics[width=\linewidth]{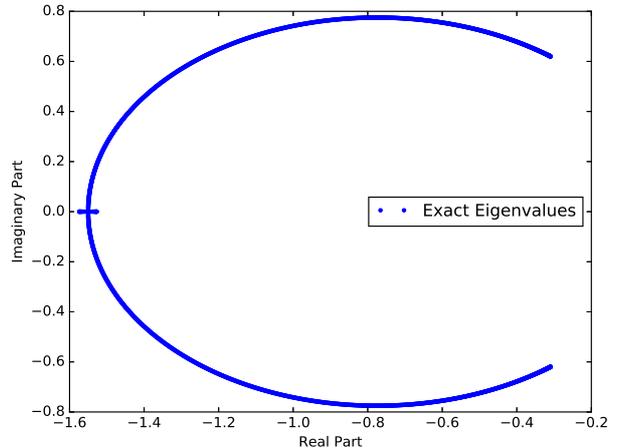}
\vspace{-0.25in}
\caption{All eigenvalues of the nonoverdamped mass-spring system for $n = 1000$, $\tau = 0.6202$ and $\kappa= 0.4807$ obtained via linearization.}
\label{fig:nonALL}
\end{center}
\end{figure}

\begin{figure}[tbh!]
\begin{center}
    \textbf{NLFEAST Eigenvalue Estimates\\ for Nonoverdamped Problem}\\
\includegraphics[width=\linewidth]{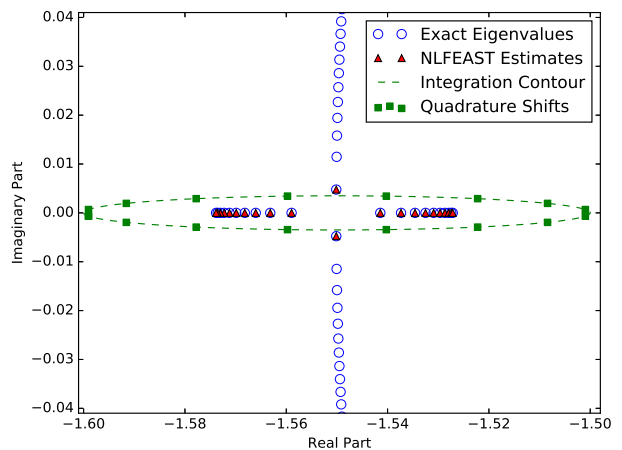}
\vspace{-0.25in}
\caption{Zoom towards all the real eigenvalues of the nonoverdamped mass-spring system for $\tau = 0.6202$, $\kappa= 0.4807$. The NLFEAST search contour is shown as a green dashed line, with the locations of the quadrature shifts indicated by green squares. Exact solutions to the nonlinear eigenvalue problem, obtained by linearization, are shown as blue circles, and the approximations obtained by using NLFEAST are indicated by red triangles. Using a subspace of dimension $m_0=22$, NLEAST correctly identifies the 20 eigenvalues inside of the search interval, in addition to the two eigenvalues that are closest to the contour without being inside of it. }
\label{fig:nonReal}
\end{center}
\end{figure}

\begin{table}[tbh!]

\begin{center}
      \textbf{Eigenvalues and Residuals for \\ the Nonoverdamped Problem}\\
  \begin{tabular}{ l || c || c | c }
    \hline\hline
         \multirow{2}{*}{$i$} & linearization & \multicolumn{2}{c}{nonlinear FEAST} \\ \cline{2-4}
		& {$\lambda_i$} & $\lambda_i$ &  $\|r_i\|_2$ \\ \hline
 1  & -1.5738531653 & -1.5738531653  & 1.65955e-14 \\ \hline
 2  & -1.5735377749 & -1.5735377749  & 1.48662e-14\\ \hline
 3  & -1.5730028887 & -1.5730028887  & 9.75627e-14\\ \hline
 4  & -1.5722332594 & -1.5722332594  & 7.75776e-14\\ \hline
 5  & -1.5712042310 & -1.5712042310  & 1.25406e-14\\ \hline
 6  & -1.5698768253 & -1.5698768253  & 1.31504e-14\\ \hline
 7  & -1.5681876058 & -1.5681876058  & 9.65162e-15\\ \hline
 8  & -1.5660250643 & -1.5660250643  & 1.16872e-14\\ \hline
 9  & -1.5631614676 & -1.5631614676  & 7.51754e-14\\ \hline
 10 & -1.5589513444 & -1.5589513444  & 8.23048e-14\\ \hline
 11 & -1.5414378153 & -1.5414378153  & 1.14765e-14\\ \hline
 12 & -1.5373437441 & -1.5373437441  & 1.01760e-14\\ \hline
 13 & -1.5345839864 & -1.5345839864  & 4.87860e-14\\ \hline
 14 & -1.5325130699 & -1.5325130699  & 3.06639e-14\\ \hline
 15 & -1.5309032607 & -1.5309032607  & 5.33027e-14\\ \hline
 16 & -1.5296430495 & -1.5296430495  & 1.72740e-14\\ \hline
 17 & -1.5286689994 & -1.5286689994  & 1.56610e-14\\ \hline
 18 & -1.5279421315 & -1.5279421315  & 3.50522e-14\\ \hline
 19 & -1.5274377896 & -1.5274377896  & 1.76889e-14\\ \hline
 20 & -1.5271407258 & -1.5271407258  & 1.82111e-14\\ \hline
    \hline
  \end{tabular}
\end{center}
\caption{All real eigenvalues of the nonoverdamped mass-spring system for $n=1000$, $\tau = 0.6202$ and
$\kappa= 0.4807$ within the interval $(-1.6,-1.5)$.
}
\label{tab:nonoverdamped}
\end{table}
\begin{table}[tbh!]
\begin{center}
    \textbf{NLFEAST Iterations vs. \\ contour size for constant $n_c$}\\
\begin{tabular}{ c || c | c}
\hline\hline
\multirow{2}{*}{Interval} & \multicolumn{2}{c}{$\#$ of NLFEAST} \\
\balancedVPhantom{2em}$(a,b)$ & eigenvalues & iterations \\ \hline
\hline
$(-1.6, -1.5)$ & $20$ & $3$ \\ \hline
$(-1.45, -1.65)$ & $20$ & $4$ \\ \hline
$(-1.35, -1.75)$ & $20$ & $10$ \\ \hline
\hline \hline
\end{tabular}
\end{center}
\caption{Number of eigenvalues of the nonoverdamped mass-spring system for $n=1000$, $\tau = 0.6202$ and
$\kappa= 0.4807$ found within the interval $(a,b)$. The number of NLFEAST iterations that is required for convergence increases as the interval size is increased,
because the integration quadrature rule is less accurate when the size of the contour is increased without a proportional increase in the number of quadrature nodes.}
\label{tab:interval}
\end{table}
\end{example}

%
%
%

%

%
%

\begin{example}{\it An overdamped mass-spring system}

\smallskip
\noindent
We now consider the same mass-spring system as in Example~\ref{Exp1}, but with new parameter values $\tau = 10$ and $\kappa = 5$.
This choice of parameters results in a hyperbolic QEP with real and non-positive eigenvalues~\cite[\S 3.9]{TisM01}.
For $n=50$ all $100$ eigenvalues of $Q$ obtained via linearization are plotted in Figure~\ref{fig:Exp2nonALL}. 
Half of the eigenvalues are clustered very close to zero, and the other half are distributed throughout the interval $(-50,-9)$.
The spectral gap between the lowest $n$ and the highest $n$ eigenvalues is a characteristic of this class of problems.

In the case of an overdamped system there are no oscillations; the dynamics of the system consist entirely of decreasing exponentials. 
The fastest dynamics of the system are determined by the eigenvectors whose eigenvalues come before the eigenvalue gap, i.e.,
the ones with the largest magnitudes. As an example, we calculate the real eigenvalues lying within the interval $(-30, -11)$.
The contour is chosen as a circle centered at the midpoint of the interval $(-30,-11)$ with radius $r = 19$, using $n_c = 8$ quadrature nodes, and the convergence criteria for the residual is $10^{-10}$. With the subspace of size $m_0 = 25$,
the nonlinear FEAST algorithm finds all $m = 19$ eigenvalues within the interval in ten ($10$) FEAST iterations.

Figure \ref{fig:Exp2Feast} illustrates the the eigenvalues of the problem, the FEAST contour and quadrature points, and the resulting eigenvalue estimates that are calculated by NLFEAST.

\begin{figure}[tbh!]
\begin{center}
    \textbf{All Eigenvalues of the \\Overdamped Problem}
\includegraphics[width=\linewidth]{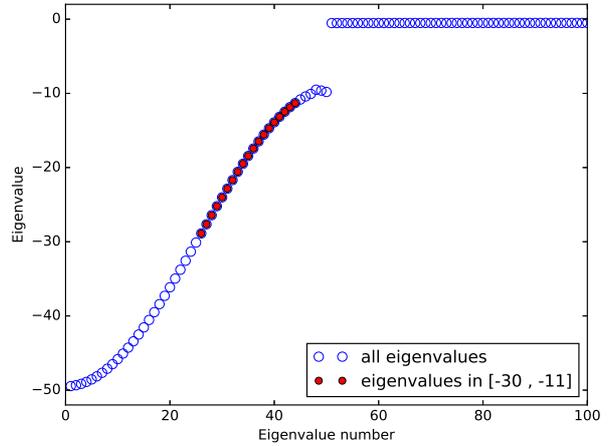}
\vspace{-0.25in}
\caption{All eigenvalues of the overdamped mass-spring system with $\tau = 10$, $\kappa= 5$ and $n = 50$ obtained via linearization. 
The eigenvalues we seek with NLFEAST are highlighted in red. A large number of eigenvalues is clustered relatively close to zero; these are separated from the rest by a spectral gap.}
\label{fig:Exp2nonALL}
\end{center}
\end{figure}

\begin{figure}[tbh!]
\begin{center}
    \textbf{NLFEAST Eigenvalue Estimates\\ for Overdamped Problem}\\
\includegraphics[width=\linewidth]{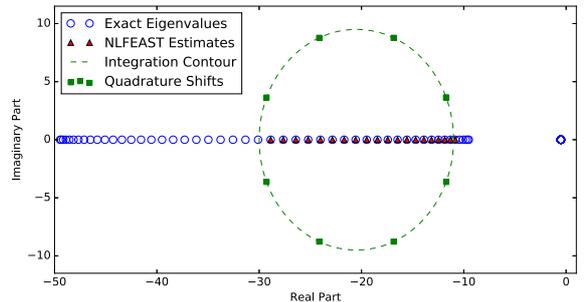}
\vspace{-0.25in}
\caption{The nonlinear FEAST approximations of $m = 19$ eigenvalues inside interval $(-30,-11)$ of the overdamped mass-spring system
for $n= 50$, $\tau = 10$ and $\kappa= 5$.}
\label{fig:Exp2Feast}
\end{center}
\end{figure}
\end{example}

%
%
%

\begin{figure}[tbh!]
\begin{center}
    \textbf{All Eigenvalues for\\ Scattering Resonance Problem}\\
\includegraphics[width=\linewidth]{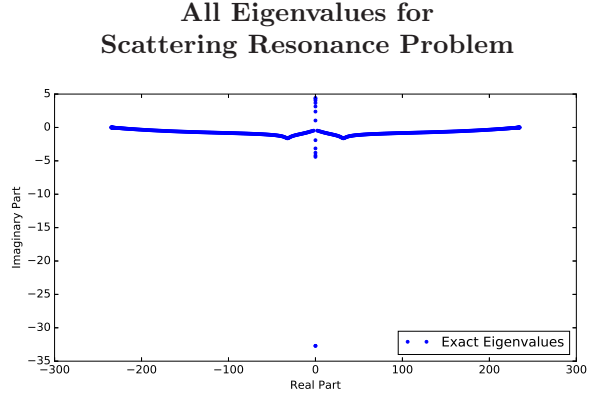}
\vspace{-0.25in}
\caption{All scattering resonances for potential $V_0 = 10$ obtained via linearization \eqref{eq:resonanceLin}.}.
\label{fig:Exp5All}
\end{center}
\end{figure}

\begin{figure}[tbh!]
\begin{center}
    \textbf{NLFEAST Eigenvalue Estimates\\ for Scattering Resonances}\\
\includegraphics[width=\linewidth]{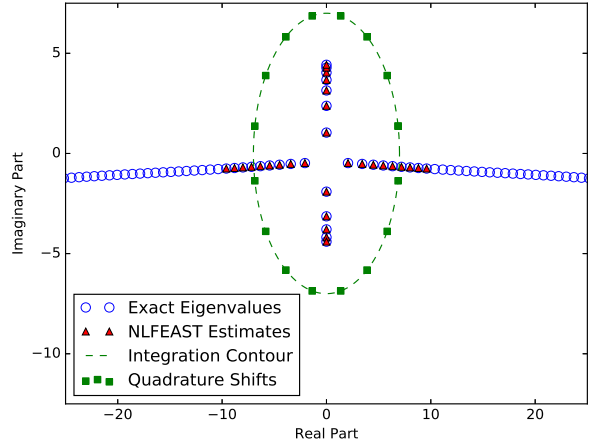}
\vspace{-0.25in}
\caption{The nonlinear FEAST approximations of $m_0=30$ eigenvalues for the scattering resonance problem for potential $V_0 = 10$. NLFEAST captures all of the 22 eigenvalues inside of the integration contour, plus the eight (8) eigenvalues that are closest to the integration contour while still being outside of it. }
\label{fig:Exp5_FEAST_15}
\end{center}
\end{figure}

\vspace{0.1in}
\begin{example}{\it Scattering resonances in 1D}

\noindent
Another application in which nonlinear eigenvalue problems arise is open boundary quantum transmission problems. With open boundary quantum transmission problems, one seeks to calculate the quasi-bound states of a quantum potential where particles can either enter or leave the system, preferably without having to explicitly model the external sources or sinks to which that quantum potential is connected. This can be done by solving a nonlinear eigenvalue problem \cite{ShaPL95}, in contrast to many other problems in quantum mechanics that can be approached by solving a linear eigenvalue problem with the system Hamiltonian.

As an example of open boundary quantum transmission problems, let us consider the problem of scattering resonances in $1$D with the following compactly supported finite square model potential
\begin{equation}
\label{eq:potential}
V(r) = \begin{cases}
             -V_0, \quad r \in [-L,L] \\
              \hspace{0.12in}  0, \hspace{0.2in} \mbox{otherwise}
            \end{cases},
\end{equation}
with $V_0 > 0$ and width $2L = \pi\sqrt{2}$~\cite{CanN17}.
%

We are interested to determine the \textit{Siegert states}~\cite{Sie39} $u \in H^{1}(-L,L)$ and the associated resonances $k \in \C$ such that for all $v \in H^{1}(-L,L)$
\begin{equation}
\begin{aligned}
\label{eq:resonanceVF}
\int\limits_{-L}^{L} u'v' + V u\bar{u}' dx  & =   \imath k \big( u(L)\bar v(L) + u(-L)\bar v(-L)\big) \\
					    & + \ k^2\int\limits_{-L}^{L}u\bar v dx.
\end{aligned}
\end{equation}
\noindent
In the case of $V: \R \rightarrow \C$ being bounded with compact support in $[-L,L]$, the set of discrete solutions
$(u_{i}, k_{i}) \in H^{1}(-R,R) \times \C$ of \eqref{eq:resonanceVF} can be approximated using the finite element space $V_h \subset H^{1}(-L,L)$
associated with the grid $x_j = -L + jh, j=0,\ldots, n+1$, $h = \frac{2L}{n+1}$. This discretization approach results in the following quadratic eigenvalue problem

\begin{equation}
\label{eq:resonanceQEP}
T(k_h)u_h = \Big(k_h^2A_h + \imath k_h B_h - C_h \Big)u_h = 0,
\end{equation}
where
\[
A_h = \frac{h}{6}\begin{bmatrix}
       2 & 1 & 0 & 0 & \ldots & 0 \\
       1 & 4 & 1 & 0 & \ldots & 0 \\
       0 & 1 & 4 & 1 & \ldots & 0 \\
  \vdots & \vdots & \vdots & \ddots & \vdots & \vdots\\
  0 & \ldots & 0 & 1 & 4 & 1 \\
  0 & \ldots & 0 & 0 & 1 & 2
      \end{bmatrix} \ \in \ \R^{n+2 \times n+2},
\]
\[  B_h = \begin{bmatrix}
			    1 & 0 & \ldots & 0 & 0 \\
			    0 & 0 & \ldots & 0 & 0 \\
			    \vdots & \vdots & \vdots & \vdots & \vdots\\
			    0 & 0 & \ldots & 0 & 0 \\
			    0 & 0 & \ldots & 0 & 1
			   \end{bmatrix} \ \in \ \R^{n+2 \times n+2},
\]
and
\[
			   C_h = \frac{1}{h} \begin{bmatrix}
									 1 & -1 & 0 & 0 & \ldots & 0 \\
									-1 & 2 & -1 & 0 & \ldots & 0 \\
									0 & -1 & 2 & -1 & \ldots & 0 \\
									\vdots & \vdots & \vdots & \ddots & \vdots & \vdots \\
									0 & \ldots & 0 & -1 & 2 & -1 \\
									0 & \ldots & 0 & 0 & -1 & 1
									\end{bmatrix}  + V_0 A_h \ \in \ \R^{n+2 \times n+2}.
\]

\noindent
The associated linear eigenvalue problem has a form
\begin{equation}
\label{eq:resonanceLin}
\begin{bmatrix}
-\imath B_h & C_h\\
 I  & O
\end{bmatrix} \begin{bmatrix}
		k_h u_h \\
		u_h
	      \end{bmatrix} = k_h \begin{bmatrix}
				      A_h & O\\
				      O & I
				      \end{bmatrix} \begin{bmatrix}
						    k_h u_h\\
						    u_h
						    \end{bmatrix}.
\end{equation}
\end{example}
\noindent
A finite element discretization of \eqref{eq:resonanceVF} over the grid of $n = 302$ points results
in a quadratic eigenvalue problem of size $302$ with scattering resonances plotted in Figure~\ref{fig:Exp5All}.
We are interested in all twenty-two $(22)$ complex scattering resonances lying inside the circle centered at $0.0$ with
radius $r = 15.5$, see Figure~\ref{fig:Exp5All}. The nonlinear FEAST computes $10^{-10}$ accurate approximations of those
twenty-two ($22$) scattering resonances in four $(4)$ iterations using $n_c = 16$ integration nodes and the subspace of size $m_0 = 30$, see Figure~\ref{fig:Exp5_FEAST_15}.

%

%
%
%
%

\begin{example}{\it Quartic Eigenvalue Problem.}
As an example of a polynomial eigenvalue problem with degree larger than two, let us consider the following quartic problem
$$
P(\lambda)x  = (\lambda^4A_4 + \lambda^3A_3 + \lambda^2 A_2 + \lambda A_1 + A_0)x = 0.
$$
Eigenvalue problems of this form can come from, for example, discretizations of the Orr-Sommerfeld equation~\cite{BriM84,TisH01}. The Orr-Sommerfeld equation arises from a linearization of the incompressible Navier-Stokes equation in which the perturbations of the pressure and velocity  are assumed to be periodic in time.

To illustrate the behavior of the nonlinear FEAST algorithm, we use a simple example of a quartic eigenvalue problem provided by the NLEVP collection \cite{nlevp_collection}: 
the so-called \textit{butterfly} problem (distribution of eigenvalues in the complex plane resembles a shape of a butterfly).
The \textit{butterfly} problem is a $64 \times 64$ structured quartic matrix pencil with $256$ eigenvalues, the construction of which is described in \cite{MehW02}.
We use the NLFEAST algorithm to calculate the eigenvalues that are located inside of some arbitrarily chosen region $\mathcal{C}$ in the complex plane. This problem is illustrated in Figure \ref{fig:Ex5_problem}.

\begin{figure}[tbh!]
\begin{center}
  \textbf{Eigenvalues of the Butterfly Problem}
\includegraphics[width=\linewidth]{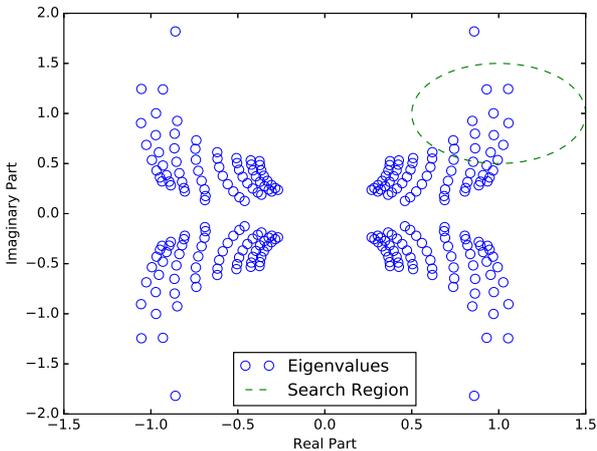}
\vspace{-0.25in}
\caption{An illustration of the locations of the eigenvalues of the \textit{butterfly} problem in the complex plane, as well as the search region $\mathcal{C}$ in which we calculate eigenvalues using NLFEAST.}
\label{fig:Ex5_problem}
\end{center}
\end{figure}

We calculate 13 eigenvalues inside of the indicated region by using a subspace of dimension $m_0=15$, using several different 
numbers of quadrature nodes $n_c$. The largest (at each iteration) eigenvector residual associated with the eigenvectors whose eigenvalues are inside the search region $\mathcal{C}$
is plotted in Figure \ref{fig:Ex5_CP}. Using $n_c=8$ quadrature nodes,
NLFEAST does not converge at all. We can achieve steady convergence by using $n_c = 32$ quadrature nodes, and the rate of convergence increases with increasing values of $n_c$. 
For $n_c=128$, convergence to the desired tolerance of $10^{-10}$ occurs in only five $(5)$ NLFEAST iterations. 
Because the linear system for each individual quadrature node is independent of the linear systems associated with all the other quadrature nodes,
the rate of convergence of NLFEAST can be systematically improved by using additional parallel processing power to solve a larger number of linear systems simultaneously in parallel.

\begin{figure}[tbh!]
\begin{center}
  \textbf{Convergence of NLFEAST for Butterfly Problem}
\includegraphics[width=\linewidth]{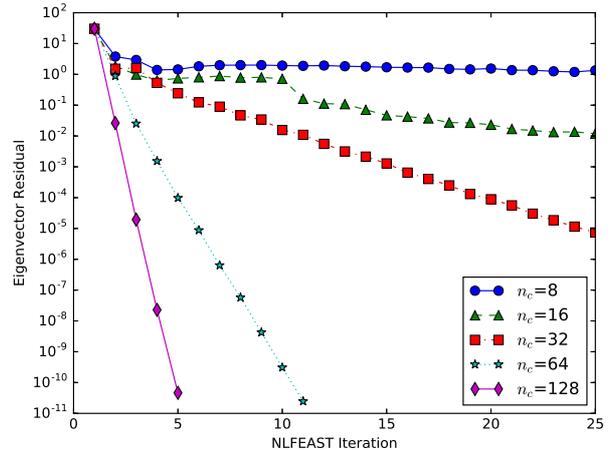}
\vspace{-0.25in}
\caption{Convergence of the NLFEAST algorithm for the eigenvectors whose eigenvalues are inside the search region when solving the {\em butterfly} problem. Convergence trajectories are shown for different numbers of integration quadratute points $n_c$. NLFEAST is not able to converge effectively when using only 8 quadrature points, and converges rapidly when using 128 points. When using enough parallel processing power, a single iteration takes the same amount of time regardless of the number of quadrature points that are used.}
\label{fig:Ex5_CP}
\end{center}
\end{figure}

Figure \ref{fig:Ex5_cp32vs8} 
shows the NLFEAST-estimated eigenvalues for the experiments from Figure \ref{fig:Ex5_CP} that use $n_c=8$ and $n_c=32$ quadrature points in the numerical integration. 
The $n_c=8$ case is not able to converge because the integration is not sufficiently accurate to achieve convergence of the two eigenvalues 
that are farthest-separated from the main cluster of eigenvalues; using $n_c=32$ allows the accurate convergence of NLFEAST for all of eigenvalues inside the search region $\mathcal{C}$.

\begin{figure}[tbh!]
\begin{center}
  \textbf{NLFEAST Eigenvalue Estimates\\ for Butterfly Problem}\\
\includegraphics[width=\linewidth]{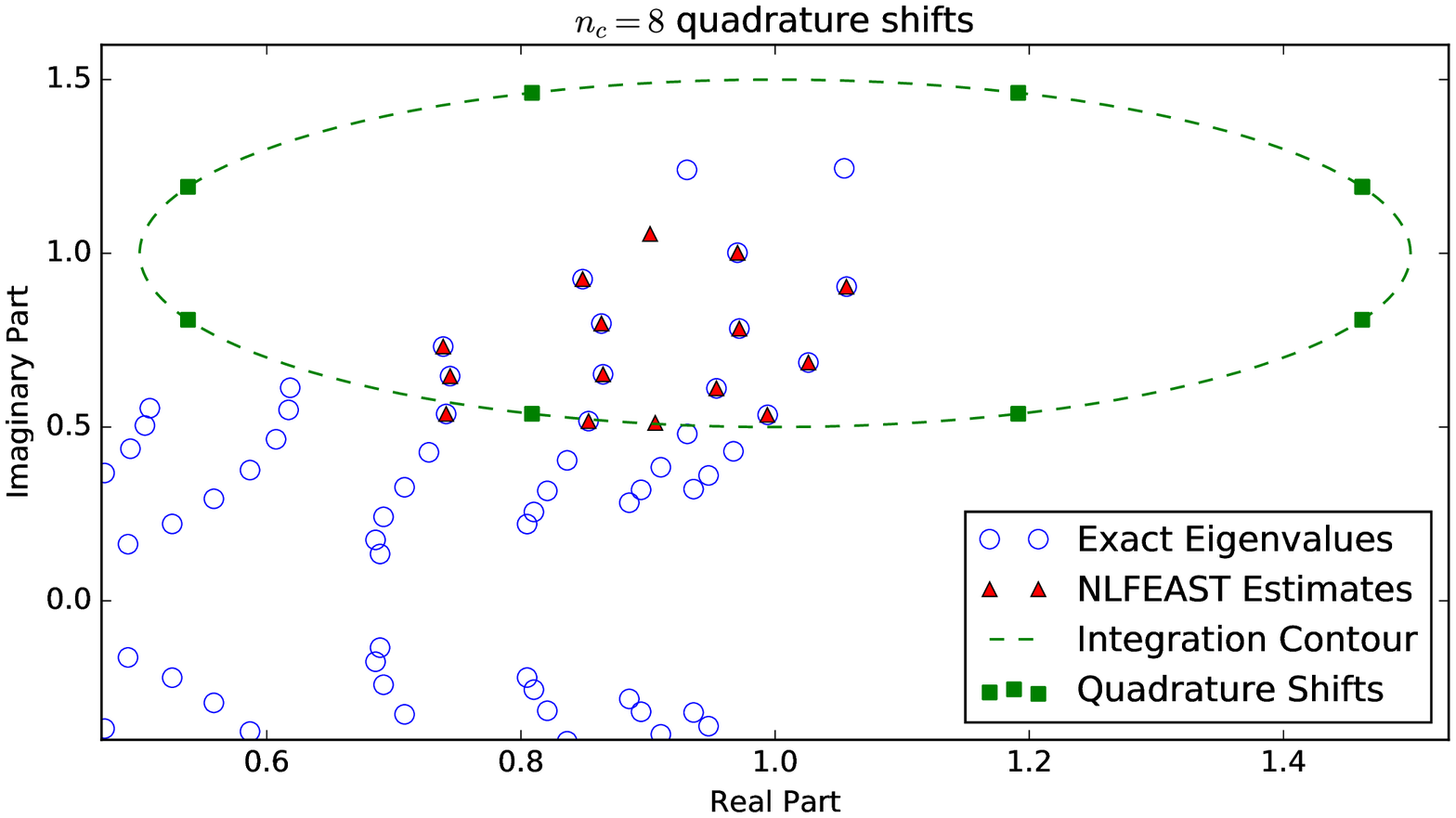}\\
\includegraphics[width=\linewidth]{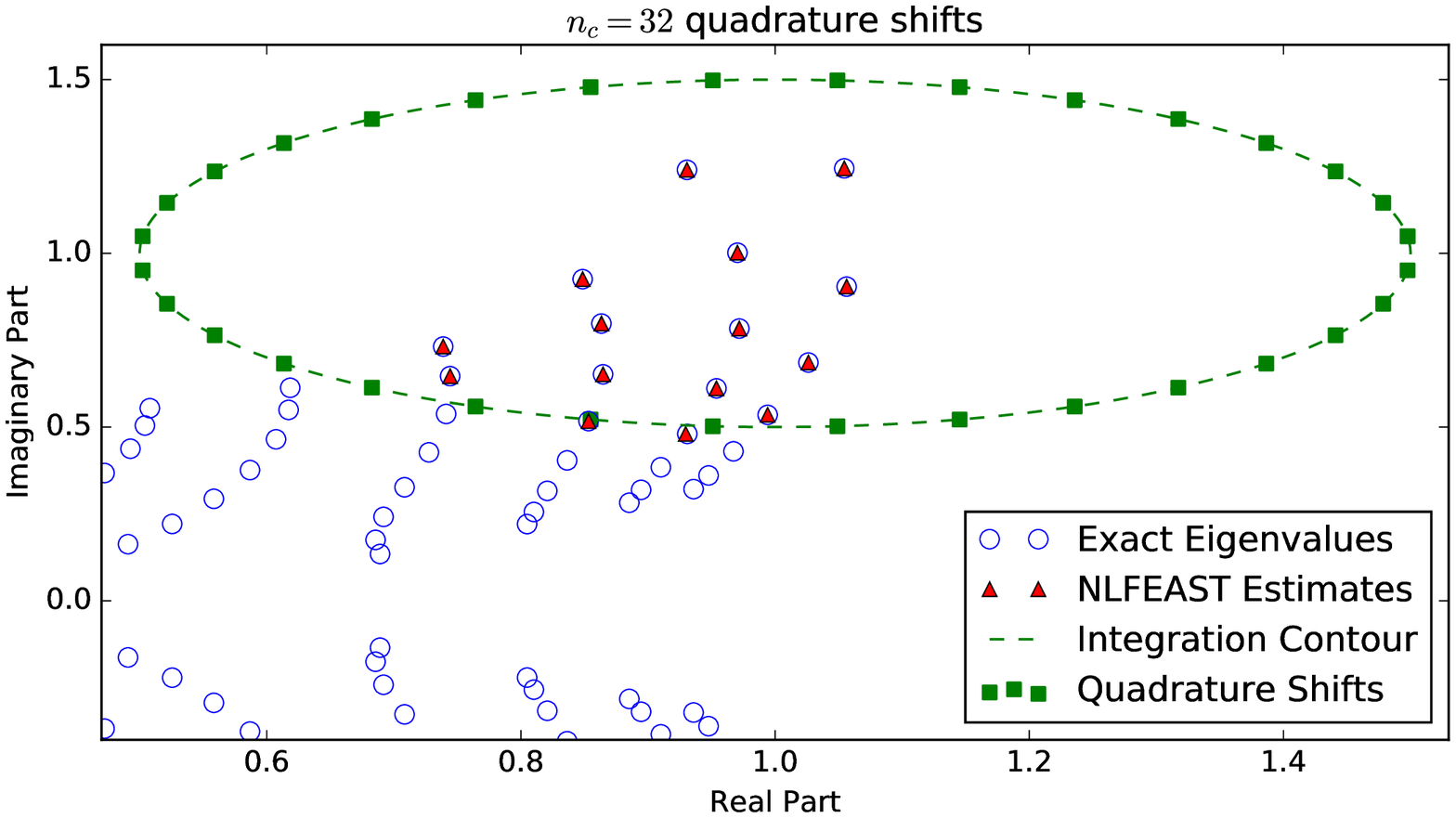}
\vspace{-0.25in}
\caption{Illustrations of the NLFEAST-estimated eigenvalues for the {\em butterfly} problem. 
The results when using $n_c=8$ quadrature nodes (top), and the results when using $n_c=32$ quadrature nodes (bottom).
Increasing the number of quadrature points improves the quality of the integration, allowing NLFEAST to better-identify eigenvalues that are farther away from the main cluster.}
\label{fig:Ex5_cp32vs8}
\end{center}
\end{figure}
 \end{example}

\section{Summary and conclusions}

In this paper we have described an extension of the linear FEAST eigenvalue algorithm for solving nonlinear eigenvalue problems. 
The resulting nonlinear FEAST algorithm (NLFEAST) uses the same series of operations as the linear FEAST algorithm,
but with a modified contour integral that allows for using a fixed collection of shifts and a fixed subspace dimension in solving nonlinear eigenvalue problems. Where the linear FEAST algorithm can be interpretted as a generalization of shift-and-invert iterations that can use multiple shifts, the nonlinear FEAST algorithm can be interpretted as a generalization of residual inverse iterations that can use multiple shifts.

Like the linear version, the NLFEAST algorithm can be used to calculate eigenvectors corresponding to the eigenvalues located in a specific, user-defined region in the complex plane,
allowing for the parallel calculation of large numbers of eigenpairs. Moreover, analogously to the linear FEAST, the convergence rate of the NLFEAST can be systematically improved by solving
additional linear systems in parallel to refine the numerical contour integration.

In this paper, for the sake of simplicity, we have treated only polynomial eigenvalue problems.
This makes the implementation of NLFEAST relatively straight-forward, in the sense that the reduced, nonlinear eigenvalue problem in the NLFEAST algorithm can be solved easily via linearization.
We expect, however, that the NLFEAST algorithm as described in this paper will prove to be a general method of solution for nonlinear eigenvalue problems of any form, rather than just polynomial eigenvalue problems.
In the case of more general (non-polynomial) nonlinear eigenvalue problems, the reduced nonlinear eigenvalue problem will be itself of an arbitrary form, and therefore will require using
a Newton- or a projection-type solution method instead of a simple linearization technique.

\section*{Acknowledgements}

The authors would like to thank Eric Cances and Boris Nectoux for providing the scattering problem example. The work of Agnieszka Miedlar was
partially supported by the Simons Foundation Collaboration Grant for Mathematicians \textit{Eigenvalue Computations in Modern Applications}. This work was also supported by the National Science Foundation, under grant \#CCF-1510010.

\section*{References}

\bibliography{mybibfile}

\end{document}